\documentclass[conference]{llncs}

\newcommand{\etal}{{\em et al.}}
\newcommand{\ignore}[1]{}
\usepackage{epsfig}
\usepackage{graphicx}
\usepackage{subfig}

\newcommand{\figref}[1]{Fig.~\ref{#1}}

\newcommand{\secref}[1]{Sec.~\ref{#1}}
\newcommand{\eqref}[1]{Eq.~(\ref{#1})}

\begin{document}
\mainmatter  
\title{On the Dynamics of IP Address Allocation and Availability of End-Hosts}

\author{
Oded Argon$^1$ \and Anat Bremler-Barr$^2$ \and Osnat Mokryn$^3$ \and Dvir Schirman$^1$ \\ Yuval Shavitt$^1$ \and Udi Weinsberg$^1$}
\institute{School of Electrical Engineering Tel-Aviv University, Israel \and
Computer Science Dept. Interdisciplinary Center, Herzliya, Israel \and School of Computer Science, Tel Aviv-Yaffo College, Israel}

\maketitle

\begin{abstract} The availability of end-hosts and their assigned routable IP addresses has impact on the
ability to fight spammers and attackers, and on peer-to-peer application performance.  Previous
works study the availability of hosts mostly by using either active pinging or by studying access to a mail
service, both approaches suffer from inherent inaccuracies. We take a different approach by measuring the IP
addresses periodically reported by a uniquely identified group of the hosts running the DIMES agent. This fresh
approach provides a chance to measure the true availability of end-hosts and the dynamics of their assigned routable
IP addresses. Using a two month study of 1804 hosts, we find that over 60\% of the hosts have a fixed IP address
and 90\% median availability, while some of the remaining hosts have more than 30 different IPs. For those that  
have periodically changing IP addresses, we find that the median average period per AS is roughly 24 hours,
with a strong relation between the offline time and the probability of
altering IP address.
\end{abstract}

\section{Introduction}


Many ISPs dynamically assign IPv4 addresses to hosts, using
methods such as Dynamic Host Configuration Protocol (DHCP) and
Internet Protocol Control Protocol (IPCP).
\ignore{
IPCP simply provides an IP address
to a customer from
a pool of IP addresses maintained by the ISP.
DHCP servers attach a \emph{lease time} to each assigned address, which, upon
expiration, must be renewed by the customer. The DHCP server can either
accept the renewal request, or refuse it, causing the customer to request
a new IP address.
}
Both methods enable ISPs to change the routable IP addresses assigned to customers, 
either when a \emph{lease time} expires (in DHCP) or on customer modem restart (in IPCP).

\ignore{
Since ISPs often sell premium services that provide a fixed IP for a paying customer,
they have an incentive to replace the IP address for non-premium customers on
each modem restart or lease expiration. Additionally, deploying DHCP
servers is quite common, even when there is a sufficient number of routable
IP addresses, since it makes network management significantly easier, by removing
the need to manually reconfigure individual machines.
}

Understanding the time period during which a host is online and reachable using the same IP address,
has implications on
various network applications and studies. Specifically,  tasks like
malicious host identification, network forensic analysis and other
blacklisting based approaches require tracking connected hosts
over time using their IP
addresses \cite{peng2004proactively,ramachandran2007filtering,Xie07howdynamic,zhuang2008characterizing,nick2006,wilcox10a}.
\ignore{
From a security aspect, the longer that a host is online and maintains a specific IP address,
it is easier to deploy various targeted attacks.}
The main assumption that lies behind the
highly popular blacklists is that a blacklisted machine is identified
by its routable IP address, which is completely invalid when facing dynamic IP addresses.
Additionally,
many peer-to-peer applications, such as file-sharing, voice chats, multi-player games and distributed storage \cite{conext09},
identify hosts using their routable IP addresses, and
need it to be stable for prolonged periods in order to provide better
service \cite{user2006}.

Capturing these dynamics, however, is a non-trivial task. Previous studies \cite{dischinger-2007-broadband,user2006,conext09} used
DNS to gain routable IP addresses of hosts, and then
performed long-term probing of these IP addresses. The underlying assumption that
the probing is performed on the same host fails if the IP address changes during the time
of the probe. Additionally,
there is no easy way to tell whether the ping reply came from the end-user machine, its network equipment, or even its ISP equipment (e.g., a gateway).
However, these
issues are commonly overlooked.

The most similar large scale study of the dynamics of IP addresses was performed by Xie~\etal~\cite{Xie07howdynamic}
who used a month-long Hotmail traces for finding more than 102 million dynamic IP addresses.
However, the authors used the Hotmail credentials in order to uniquely identify machines. This
is problematic since many users access their emails from many different machines.

A smaller study of a DHCP server was performed by Khadilkar \etal \cite{imc07},
who
studied the Georgia Tech LAWN (Local Area Walkup and Wireless Network)
DHCP server for 5 days. The authors showed a median
session time of 75 minutes, which is mainly attributed to the dynamics of
people in the campus and wireless networks. We take a broader look, and target various types of ISPs, and
the dynamics of availability for mostly stationary hosts.

\ignore{
However, other than intentional replacement of IP addresses for encouraging
premium SLAs,
there are no precise rules for defining a policy for IP replacement, or
methods to predict when a customer will not renew its lease.
As such, the lease time, which is a key configuration component in DHCP,
is a source of confusion for
network operators, mainly since it is not clear how to decide the right value.
Since the DHCP protocol is very chatty, one would prefer to
have a long lease time, so that network load and server congestion is kept to minimum.
On the other hand, when there is a high churn of hosts, long lease time
can reduce the availability of the IP addresses to new hosts.
}

\ignore{
This paper presents a fresh approach to this problem by leveraging
periodical announcements
 originated from a set of over 1800 end-hosts, and using these to
 measure their
 availability and IP address allocation over a period of two months.
 We introduces new tools from
 signal processing for some of the analysis and were able to ...
}

In this paper, we take a
fresh approach to study the dynamics of IP addresses, by
leveraging periodical announcements originated from a set of over 1800
end hosts, and further validate our results over a large dataset
provided to us by a commercial Internet
advertising company.  We analyze statistics of the IP address holding time intervals, and host availability.
For a better understanding of the interval dynamics, we
introduce tools from signal processing that enable us to identify
periodic behavior, when exists, and study the periodicity statistics.

Our results show that over a period of 2 months, over 60\% of the hosts
have a single IP (which we term a fixed IP) and a 90\% median online time. Hosts with alternating IPs exhibit a
much lower 40\% median online time, and some have periodic patterns of
IP alternation, with a median period of 24 hours. We also found
a strong relation between the length of offline times and the probability
that a host will change its IP when coming back online.

\vspace{-2mm}

\section{Measurement Setup}
\label{measurement-infra}
\vspace{-2mm}
\subsection{Dataset}
The data used in this paper is
obtained from DIMES \cite{dimes-ccr}, a community-based
Internet measurements system, which
utilizes hundreds of software agents installed on
user PCs and on PlanetLab servers. In a DIMES installation, a user
installs a single agent on each machine. Each agent
has a unique ID which is associated with the machine it is installed on.
%

When an agent is online, it asks for a measurement script from the DIMES central server, and it
performs between 2 (set by default) and 4 measurements per minute from this script.
Once all measurements are executed, the agents report their results to the
server, roughly every 30 to 60 minutes.

There are some exceptions to this normal agent behavior, which introduces
``noise" in the dataset. First, the time between consecutive accesses of an agent to
the central server can be different than expected, either due to special
measurement scripts of varying sizes, or due to short term network and server failures. We address this by
allowing delays of up to 3 hours
before considering an agent to be ``offline".


Second, some users duplicate agent settings into additional machines, instead of
installing new agnets. This results in lower times between consecutive
accesses to the server. Furthermore, when the agent is duplicated
in machines that use different routable IP addresses, it may appear as
if the same agent has constantly altering IPs. We identified
one such agent and removed it from the dataset.

Third, since an agent cannot be installed twice on the same machine,
some users run several virtual machines and install an agent on each
of them. Similarly, some users install agents on a set of hosts that
are behind a NAT.
In both cases, if standard installation is made, then there will be
several different agents that seem to behave exactly the same. If a
duplication of a single agent was performed, then it will appear that
this agent performs frequent accesses to the server, and we will gain finer sampling.
In case of multiple installation, we
identify agent clones by comparing their list of IP addresses, and
keep only one agent for each set that exhibit identical IPs.  

Finally, an agent can be installed on a laptop, which moves between
different locations, changing IP addresses, ASes, and possibly countries. Although
this may introduces a great deal of noise into the data, we later show that these
agents are not common, and there is an easily detectable gap of, at least, a couple of hours
between changes in location. The analysis performed is in the granularity of
``online" windows, and only when all IP addresses are in the same AS. When an
agent exhibits several ASes in the same ``online" window, we remove this window
from the analysis. We elaborate on this issue in the following section.

Two months of data are used, from
June 20th until August 20th in 2010. During this time, the DIMES server
was available 97.3\% of the time, with a
shortest server downtime of 2 hours and the longest of 13.3 hours. We later
detail how these affect the analysis and the results.


\subsection{Host IP Intervals} \label{intervals}
For each agent, the DIMES server provides us with an ordered list of the access timestamps, $T_i^j$
and the corresponding routable IP address, $IP_i^j$, where $j$ is the access (visit) counter. Considering
a set of $n$ visits for agent $i$, we
denote its set of accesses as $A_i=\left\{ \left(T_i^1, IP_i^1 \right) ... \left(T_i^n, IP_i^n \right) \right\}$.

For a given agent $i$, we build a list of intervals $I_i$, which holds a set of
consecutive time frames, such that each time frame starts with the first
appearance of an IP in $A_i$ and ends in its last consecutive appearance, namely, before a different IP
appears. Additionally, we assume that an agent has gone ``offline" and
thus end the interval if the time difference between consecutive
accesses to the server, $1<j<n,~\Delta T_i^j = T_i^{j+1} - T_i^j$ is more than a threshold gap $G$. 

\begin{figure}[tbh]
\centering
    \subfloat[Time between access]{
	\label{fig:avg-visit-time}
    \epsfig{file=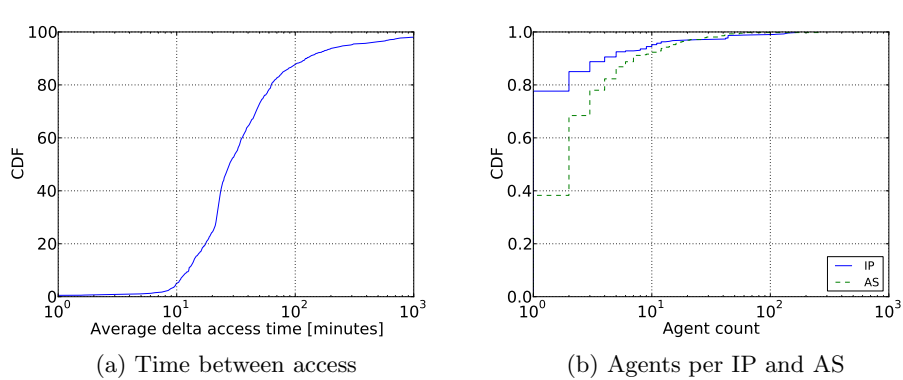,width=0.50\textwidth}
    }
    \subfloat[Agents per IP and AS]{
    \label{fig:agent-per-location}
    \epsfig{file=fig/CDF_Agent_count_per_IP_AS_tblIntervalsOfftimeAll_jun_aug_three_hours_filtering_zero_intervals,width=0.50\textwidth}
    }
    \caption{Cumulative statistics distributions, showing (a) average time between accesses of agents to the central server, and (b) number of agents per IP and AS}
    \label{fig:agent-stats}
\end{figure}

In order to determine $G$, \figref{fig:avg-visit-time} depicts the average time between consecutive accesses of each agent to the central server. As expected, 90\% of the agents have an average inter-access periods of less than 2 hours, and 
less than 10\% are due to longer scripts, offline periods and server downtimes. Being conservative, we
set $G$ to 3 hours, so that we account for the possible further variance in agent access times to the server.

The resulting $m$ intervals for a host $i$ are denoted as the set $I$, were $I_i=\left\{ \left(T_i^1, T_i^{e_1}, IP_i^1 \right)...\left(T_i^{s_m}, T_i^{e_m}, IP_i^m \right) \right\}$.
Notice that an interval set may contain the same IP address more than once, either in consecutive
intervals (in case there was an offline time between them) or in non-consecutive intervals.

Two consecutive intervals
usually have a gap between them, i.e., $\Delta T_i^j > 0$. We define an \emph{online window} as
a set of consecutive intervals that have less than $G$ time gap between them, and
an \emph{offline window} as the time between any two online windows.

The IP addresses used during gaps in an online window are unknown. Since there is no
way to determine which IP address was used and for how long,
we split the gap so that its first half 
is assigned to the interval preceding it, and the second half is assigned to the interval
following it. This simple symmetric extrapolation introduces little or no bias to the
length of the interval on average, and in any case is bounded by $G/2$, i.e., 1.5 hours.

Finally, there exist agents with intervals that have a zero length, i.e., an IP address that was
seen only once within an online window. We found that over 88\% of these intervals belong
to a single agent, while 95\% of the agents have less than 4 zero intervals. By manually
examining these zero-length intervals, we found that most of them are due to incorrect identification of the IP address,
thus we
filtered them out. In the few cases that this filtering method is incorrect, it causes the intervals
to appear longer than in reality.

\ignore{
\begin{figure}[htb]
\centering
\epsfig{file=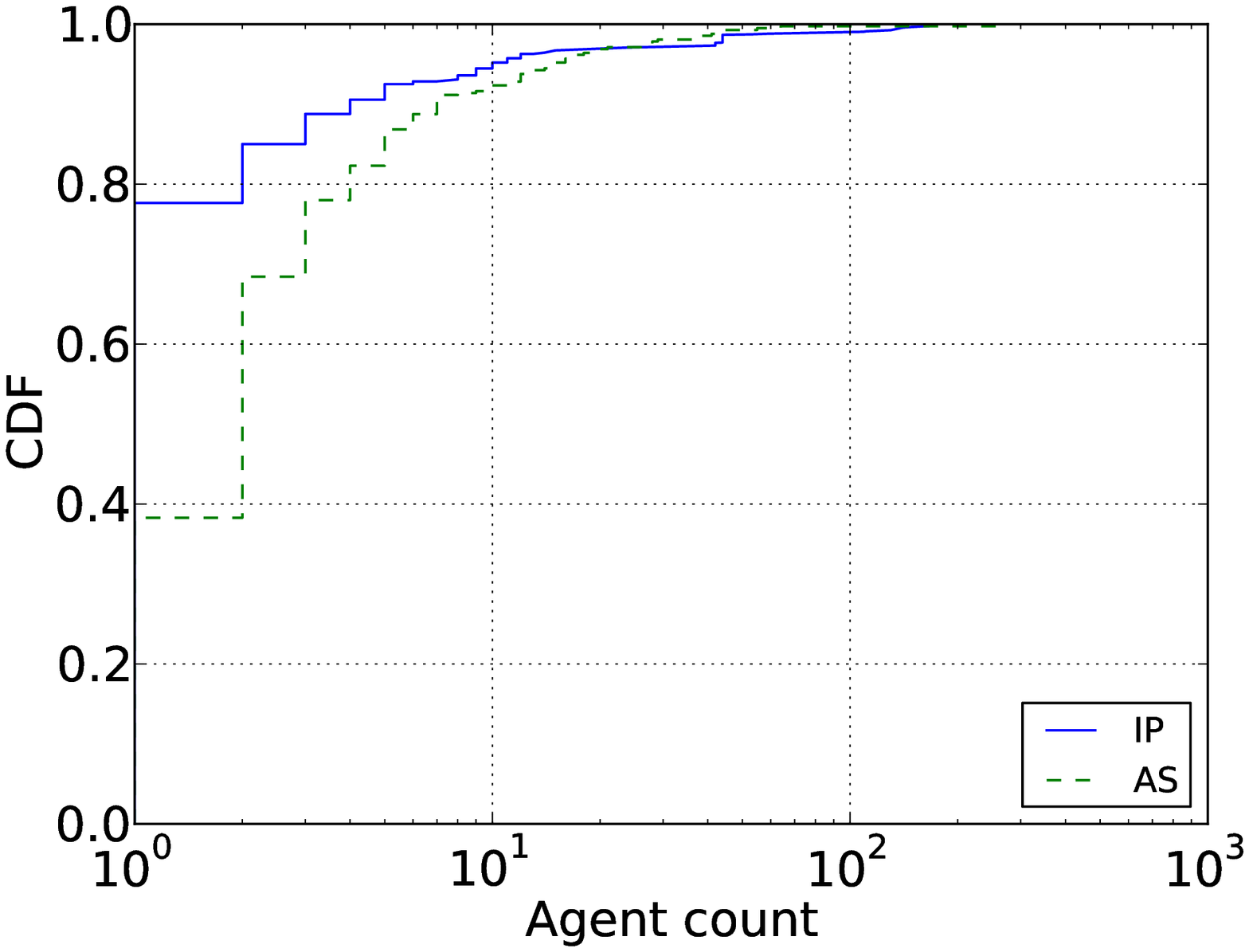,width=0.50\textwidth}
\caption{CDF of the number of agents per IP and AS}
\label{fig:agent-per-location}
\end{figure}
}

\vspace{-2mm}
\subsection{Host Statistics}
\label{agent-distribution}
Using the two month dataset provides us with roughly 8.6 million samples. Overall, 1804 agents
reported 7611 different IPs in 1037 unique address prefixes (AP) and 432 unique ASes. Agents are spread in 56 countries
in all major continents: 41.4\% are in the USA,  16\%
in Western Europe, and 15.9\% in Eastern Europe.  Some
agents are located in remote locations, such as South America (3.4\%), Africa (2 agents) and the Far East (4\%),
but these are mostly
PlanetLab servers.

Looking at the types of the ASes in our dataset using \cite{taxonomy}, we find that
27\% of the host are located in academic networks, 3\% are in tier-1
ASes, and almost 46\% are in tier-2 Ases. The remaining 24\% were not resolved, hence they
are most likely small regional or corporate ASes, as these are harder to resolve. Therefore,
the hosts in the dataset represent a wide range
of different commercial and non-commercial networks.

\ignore{
\begin{figure}[tbh]
\centering
    \subfloat[IP]{
	\label{fig:agents-per-ip}
    \epsfig{file=fig/CDF_Agent_count_per_IP_tblIntervalsOfftimeAll_jun_aug_three_hours_filtering_zero_intervals,width=0.50\textwidth}
    }
    \subfloat[AS]{
    \label{fig:agents-per-as}
    \epsfig{file=fig/CDF_Agent_count_per_AS_tblIntervalsOfftimeAll_jun_aug_three_hours_filtering_zero_intervals,width=0.50\textwidth}
    }
    \caption{CDF of the number of agents per IP and per AS}
    \label{fig:agent-per-location}
\end{figure}
}

In order to further validate that agents capture a variety of Internet locations,
\figref{fig:agent-per-location}
depicts the cumulative number of agents per IP address and AS. The figure shows that almost 80\%
of the IP addresses and 40\% of the ASes host only 1 agent. The observation that
some IPs and ASes host more than 50 agents is mainly the result of a large set of
agents behind a NAT, operated by competing groups of users of the DIMES systems, mostly located in Russia and Ukraine. In order to
account for this variance, we remove agents that exhibit the same set of IP addresses. Furthermore, when
presenting AS-level statistics, we either perform per-AS averaging or
normalization prior to the comparison.

\vspace{-8mm}
\begin{figure}[tbh]
\centering
    \subfloat[IPs per agent]{
	\label{fig:ip-per-agent}
    \epsfig{file=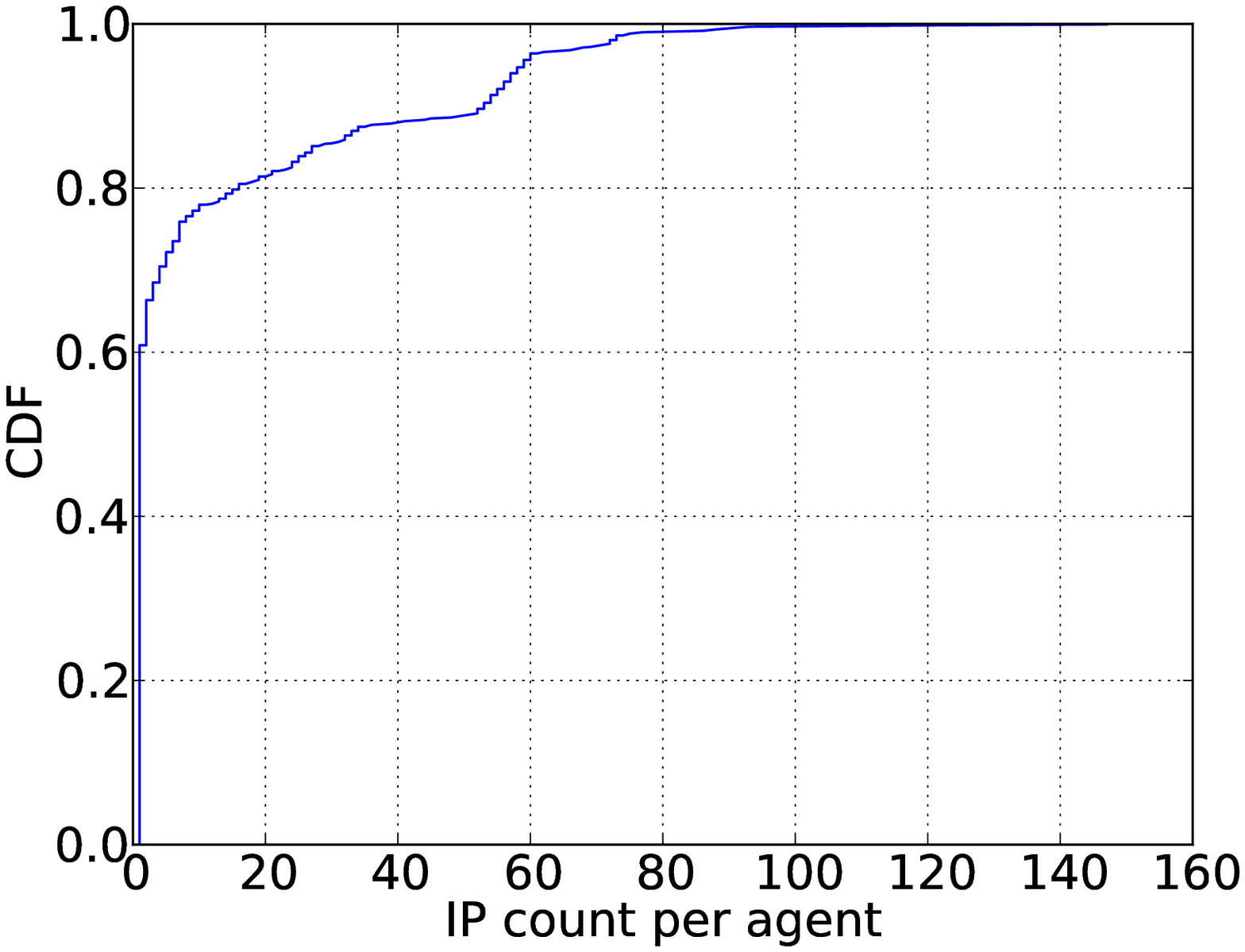,width=0.50\textwidth}
    }
    \subfloat[Fixed IP agents per AS]{
    \label{fig:static-agents}
    \epsfig{file=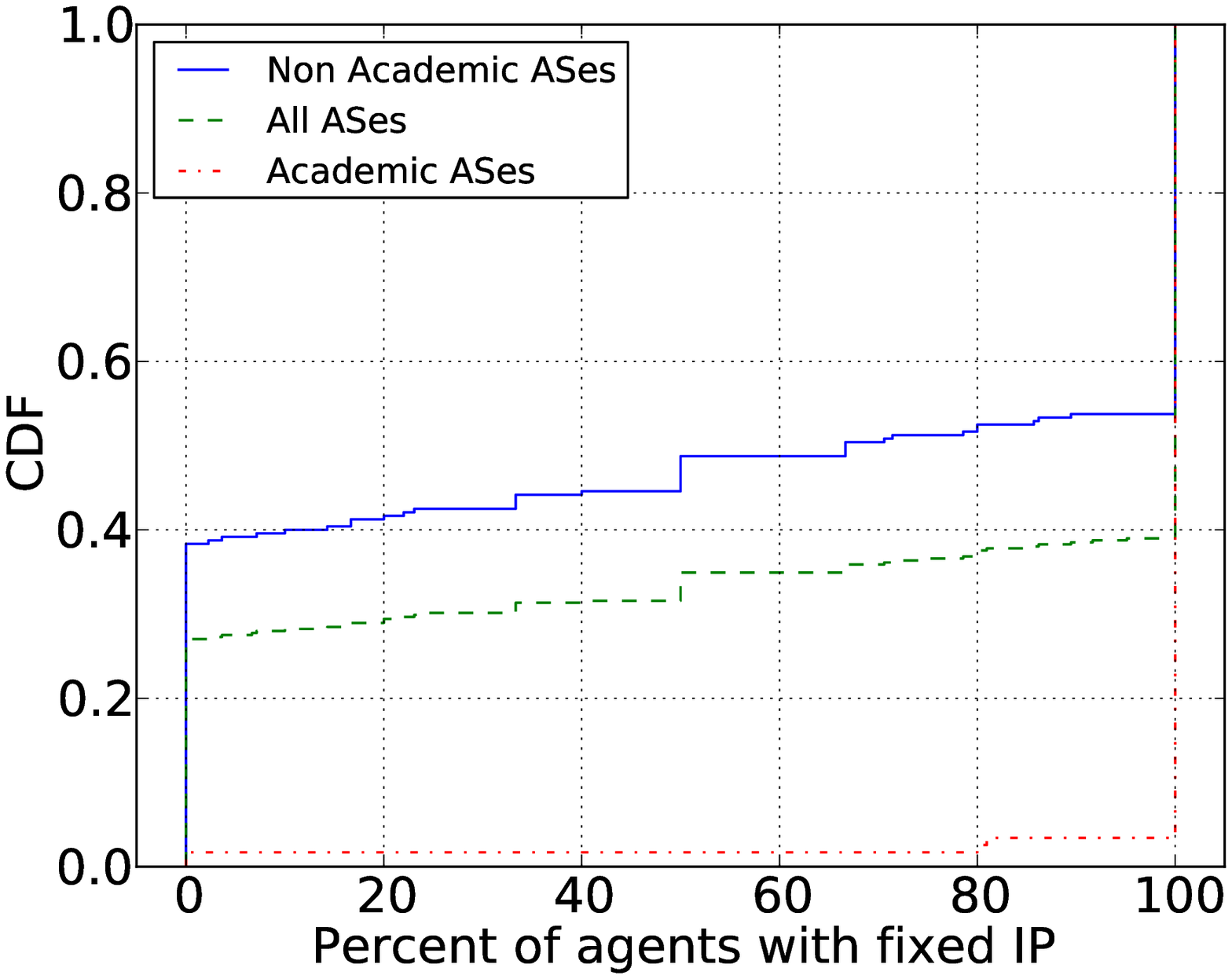,width=0.50\textwidth}
    }
    \caption{The number of IPs per agent, and the percentage of fixed agents per AS}
    \label{fig:ip-as-per-agent}
\end{figure}

\vspace{-10mm}

\subsection{IP Allocation Dynamics}
\label{static-vs-dynamic-ip}
Next, we quantify the percentage of agents that use a fixed
IP allocation, i.e., a single IP address throughout the measurement period,
and those that exhibit non-fixed IP allocation.

\figref{fig:ip-per-agent} presents the cumulative distribution of IP addresses per agent, showing
that almost 65\% of the agents have a single IP address, and
less than 5\% have 60 different IPs. We further resolved each IP address to
its corresponding AS, and found that out of the non-fixed IP agents, almost 85\% are
changing addresses within a single AS, providing a strong indication that this
is not a result of traveling with a laptop, but rather IP alternation performed by their ISPs.
In \secref{periods} we further examine the periodical patterns of
these agents.

\figref{fig:static-agents} presents the cumulative distribution of the
percentage of fixed IP agents
in each AS, showing almost a dichotomy: almost 40\% of the
non-academic ASes have no fixed IP agents, whereas almost
50\% of them have only fixed IP agents. As expected, almost all of the academic ASes have only
fixed IP agents.

The high percentage of fixed IP agents is not surprising. Valancius
\etal~\cite{conext09} showed that most users rarely turn
off their modems, hence keep their routable IP address for prolonged times, either by not
releasing it or constantly renewing their lease on time. However,
the authors also point out that energy-conscious users
switch off devices when not in need, a trend that will probably become more common
over time, and that will decrease the percentage of fixed IP hosts.

To further validate that indeed there are hosts that exhibit altering IP addresses,
and this is not an artifact solely limited to DIMES agents,
we used a large dataset provided to us by a commercial company that performs Internet advertising.
Using third party cookies, the company uniquely identifies
the click-stream of a user
in her web-browser\ignore{, and presents tailored ads}. Using a short 4-hour dataset from
the last week of Sept.\ 2010,
we analyzed over 16 million unique hosts, performing over 64 million page views. Even in this
short time-frame, almost 180k hosts (accounting for 1.1\% of the hosts) exhibit more than
one IP address, some reaching more than 10 different IP addresses. This further
strengthen our observation that IP alternations exists in many of the hosts in
today's Internet.

\begin{figure}[tbh]
\centering
    \subfloat[\% Online]{
	\label{fig:percent-online}		
	\epsfig{file=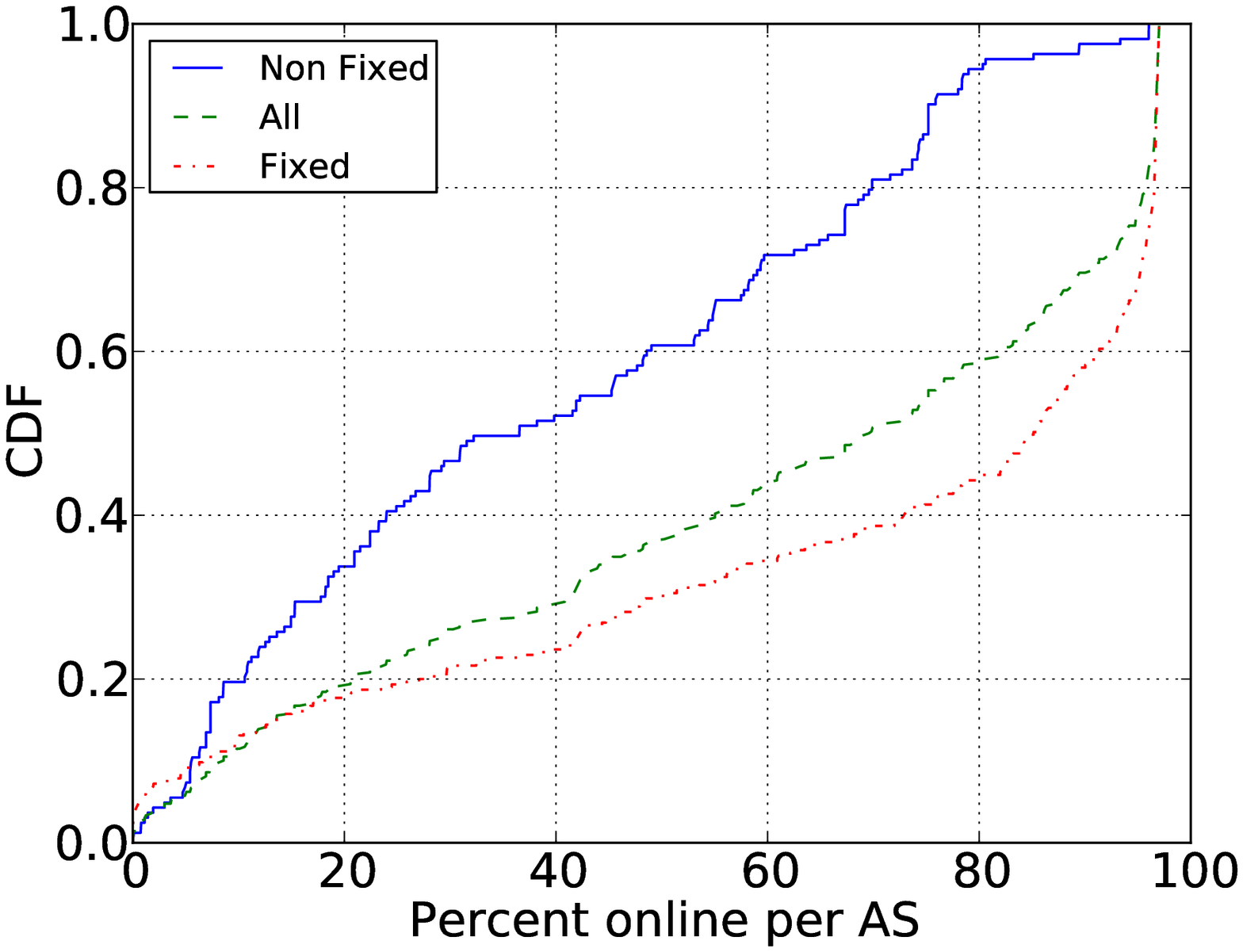,
			width=0.50\textwidth}
    }
	\subfloat[Max Interval]{
    \label{fig:max-interval}
    \epsfig{file=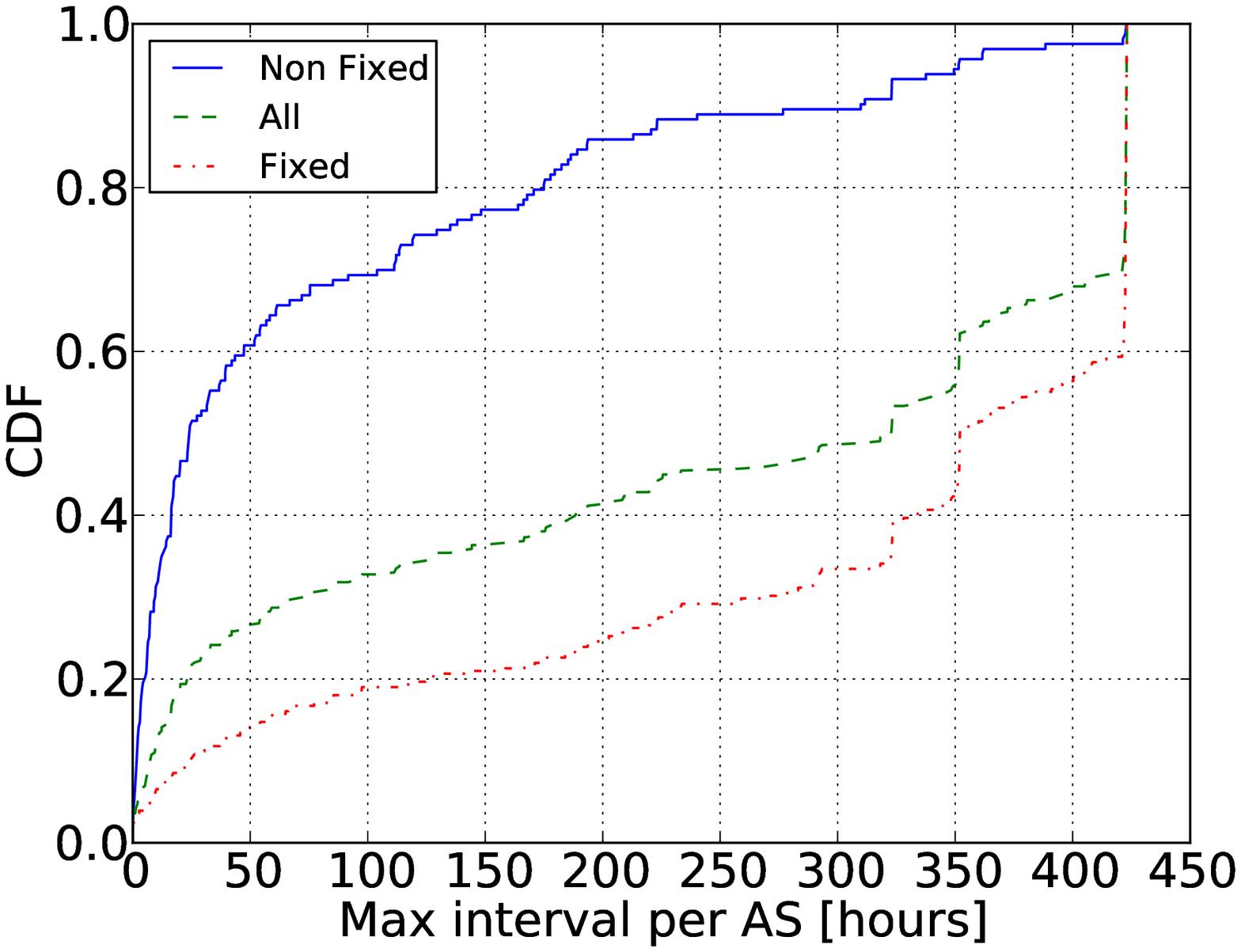,width=0.50\textwidth}
    }
    \caption{The percentage of average online time and the maximal interval length, taken over all agents, only fixed agents and only non-fixed agents in each AS}
    \label{fig:online-and-interval}
\end{figure}

\ignore{

Regardless whether the different IPs
are the result of ISP allocation policies, or host mobility, the end result is the same --
there is a large portion of hosts that have alternating IP addresses. To
further understand the availability of such hosts,
\figref{fig:online-and-interval} presents the cumulative distribution of the
overall percentage of online times, and
the maximal interval length (i.e., the maximal time that a host sustained its IP). The figure shows
that the median of static hosts exhibit 70\% online time, while the median non-static host
are not far behind with 60\% online time.

\figref{fig:max-interval}
shows that, as expected, static hosts exhibit much longer consecutive durations of sustained IP address
than non-static hosts. Interestingly, almost 30\% of the hosts
}

\vspace{-4mm}
\vspace{-2mm}

Regardless whether the different IPs
are the result of ISP allocation policies, or host mobility, the end result is the same --
there is a large portion of hosts that exhibit changing IP addresses. To
further understand the availability of hosts in the different networks, we
find, for each AS, the percentage of time it stays online. \figref{fig:percent-online}
shows a median online time of 70\% for all agents,
and a much lower 30\% for non-fixed IP agents. No AS reaches
100\% online time since the DIMES server was online roughly 97\% of this time frame,
with a maximal downtime of 13 hours, which is longer than our gap threshold of 3 hours.

These results are lower than those
previously reported \cite{conext09}, that measured an average online time of 85\% by
active probing every 15 minutes.  Their higher availability may be explained
by the probes reaching gateways, NATs, or Firewalls and not the actual hosts like in our measurements.
Alternatively, our results may be lower since the DIMES
software itself was stopped. However, this is not common, since due to the low measurement overhead,
the agents usually automatically
run in the background.

\figref{fig:max-interval} shows the cumulative distribution of the maximal interval
length (i.e., the maximal time that an agent sustained its IP). Notice that the cut-off
after 440 hours is because the DIMES server was down for 6 hours roughly 18 days after
the beginning of our experiments, hence this value is the maximal possible interval.
Although roughly 30\% of the ASes have agents that reached the maximal interval,
there exists a significant difference between the maximal interval
for fixed and non-fixed IP agents, where the latter exhibit a wide range of values,
with only 1\% that reach the maximal possible interval length.

We further checked the relation between the length of offline times and the probability to
have different IPs before and after it. Considering only non-fixed IP agents, the minimum offline time
(set to 3 hours) exhibits a probability to change IPs of 0.49, which rapidly increases to above 0.8 after
24 hours. Furthermore, there was no non-fixed IP agent that sustained its IP after an offline
time of more than 380 hours.

%
\vspace{-3mm}
\section{Inferring IP Alternation Period}
\label{periods}
\vspace{-1mm}
For each host that has more than a single IP address, we wish to identify
whether it has some periodic alternation patterns of IP addresses. Such periodicity
can be either attributed to the IP lease times or alternatively to the behavior of
the user, e.g., shutdown patterns. Since the data is quite noisy, we use signal processing
methods for inferring such periodic patterns, by constructing a time-domain signal
of IP alternation, move it to the frequency domain using Discrete Fourier Transform,
and extract the dominant frequency.

Constructing the signal is performed at the online window granularity. For each
online window, denote by IP$(t)$ the IP address it has during time $t$,
the IP alternation signal $X(n)$ is:
\begin{equation}
X(n) = \left\{
\begin{array}{l l}
  X$(n-1)$ & \quad \mbox{if IP$(nT)$ = IP$((n-1)T)$}\\
  -X$(n-1)$ & \quad \mbox{if IP$(nT)$ $\neq$ IP$((n-1)T)$}\\
  \end{array} \right.
       ~~~n=1..N
\end{equation}

\noindent where $X(0)=1$. We construct the signal $X(n)$ so that
it is discrete, by sampling the IP intervals every minute (T=1$min$). Using
one minute sampling is sufficiently fine-grained for capturing any meaningful
alternation periods, while keeping low computation overhead.

We then use Discrete Fourier Transform (DFT) over the time-domain signal,
for converting it to the frequency domain:
\ignore{
\begin{equation}
\begin{array}{l}
Y(k)=\displaystyle \sum_{n=1}^{N} X(n) \omega_N^{(n-1)(k-1)} \\
\\
\omega_N=e^{\left(2\pi i\right)/N}\\
\end{array}
\end{equation}
}

\begin{equation}
Y(k)=\displaystyle \sum_{n=1}^{N} X(n) \omega_N^{(n-1)(k-1)}~~~
,\omega_N=e^{\left(2\pi i\right)/N}
\end{equation}

In case our signal is periodic and {\em symmetric}, the frequency
matching the highest amplitude
is most likely to capture its period. Hence, for a given
frequency-domain signal, $Y(k)$, the candidate frequency, $F$,
is calculated using:
\begin{equation}
F = \displaystyle \arg\max_k \left\{ \left| Y(k) \right| \right\}
\end{equation}
The candidate period $P$ is then calculated by $P = \displaystyle 1/(2F)$.

\vspace{-4mm}

\begin{figure}[tbh]
\centering
    \subfloat[Periodic Signal]{
	\label{fig:periodic}		
	\epsfig{file=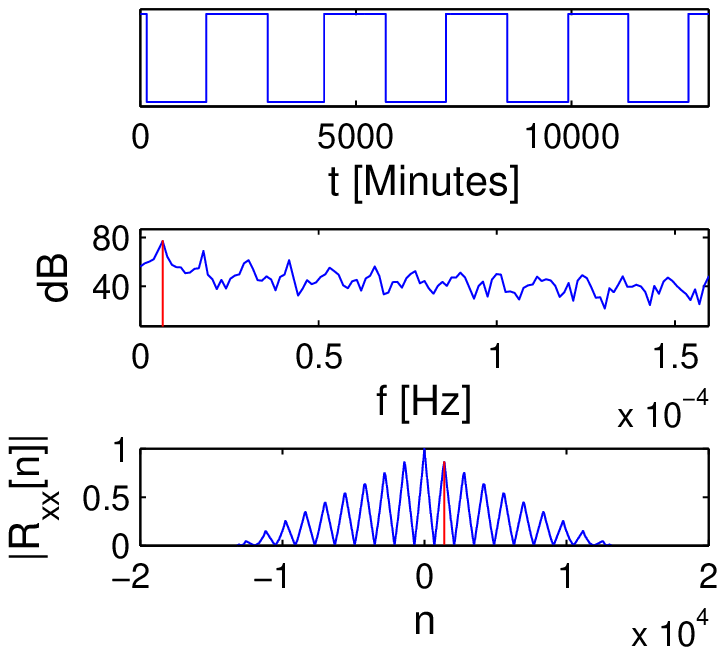,width=0.50\textwidth}
    }
    \hspace{-8mm}
    \subfloat[Non Periodic Signal]{
    \label{fig:non-periodic}
    \epsfig{file=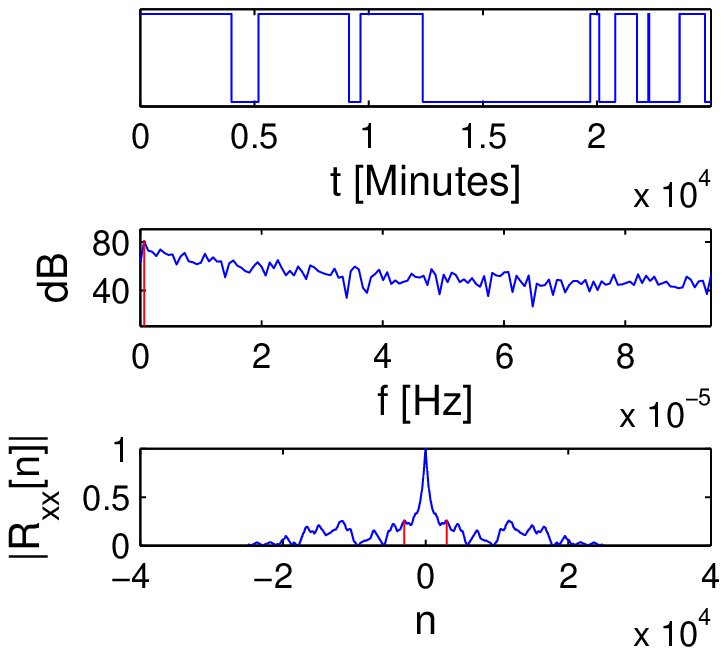,width=0.50\textwidth}
    }
    \caption{Samples of periodic and non periodic signals}
    \label{fig:periodic-and-non-periodic}
\end{figure}

\vspace{-4mm}

Several complexities arise when applying DFT to our dataset. First, square signals
result in a group of frequencies in the spectral domain (the main frequency and its odd harmonics),
lowering the amplitude of the main frequency.
Second, DFT is extremely useful for signals with amplitude noise.
However, in our analysis, the amplitude
is fixed, and the noise resides in the phase of the signal, due to the inability to
determine the exact time of IP alternation. Hence, instead of having a constant noise floor, the noise floor of the spectrum depends on the frequency, making it harder to infer the main frequency. Third, extracting $P$ by
halving the frequency assumes a duty-cycle of 50\%. This seems plausible, as
there is not real incentive for having a non-constant IP alternation policy.

The methods we use extract a $P$ for every signal,
even if it is not periodic. Therefore, we calculate
a confidence value, $\xi$, that quantifies how close the signal
is to really being periodic.
%
We do that by performing auto-correlation of the signal $X(n)$:
\begin{equation}
R_{XX}(n) = \displaystyle \frac{\sum_{m=1}^{N} X(m)X(m-n)}{N}
\end{equation}
The normalized auto-correlation is equal to 1 for any signal at the
first lag. Periodic signals exhibit
high peaks for each multiplication of the period, while for sporadic signals,
the auto-correlation peaks are expected to be much lower, and spread across many lags.

\ignore{
First we assume the signal is periodic with period $P$, and build an hypothesis signal $\widehat{X}(n)$, which is a perfect periodic signal with period $P$:
\begin{equation}
\widehat{X}(n)= \left\{
\begin{array}{l l}
  1 & \quad \mbox{if~~$2NP \leq n < (2N+1)P$}\\
  -1 & \quad \mbox{if~~$(2N+1)P \leq n < 2NP$}\\
  \end{array} \right.
  ~~~n=1..2N
\end{equation}
}

\ignore{

\begin{equation}
\widehat{X}(n)= \left\{
\begin{array}{l l}
  1 & \quad \mbox{if~~$n \leq P$}\\
  -1 & \quad \mbox{if~~$P < n$}\\
  \end{array} \right.
  ~~~n=1..2P
\end{equation}
when
\begin{equation}
\widehat{X}(n+2P)=\widehat{X}(n)
\end{equation}
}

\ignore{
\noindent We then calculate the normalized cross-correlation of $X(n)$ and $\widehat{X}(n)$:
\begin{equation}
R_{\widehat{X}X}(n) = \displaystyle \frac{\sum_{m=1}^{2N} \widehat{X}(m)X(m+n)}{N}
\end{equation}
The normalized cross-correlation is equal to 1 when $\widehat{X}(m)$ and $X(m+n)$ are exactly the same, and reduces towards 0 as the signal
increasingly deviates from the hypothesis period $P$. We build $\widehat{X}$ with $2N$ samples in order to correlate the entire length of $X$ with $\widehat{X}$ for each $n$.
}

Finally, the confidence $\xi$ is calculate as the first (smallest index) peak of the auto-correlation signal, over all lags other than the zero lag:
\ignore{
\begin{equation}
\xi = \left| R_{XX}(i) \right|,~s.t.~~i=\displaystyle \arg\min_{n \neq 0} \left\lbrace \frac{dR_{XX}}{dn}(n)=0 \right\rbrace
\end{equation}
}
\[
\xi = \left| R_{XX}(i) \right|,~s.t.~~i=\displaystyle \arg\min_{n > 0} \left\lbrace R_{XX}(n) > R_{XX}(n-1) \& R_{XX}(n) > R_{XX}(n+1)\right\rbrace 
\]

\figref{fig:periodic-and-non-periodic} presents two samples of IP alternation signals, their DFTs and $R_{XX}$. The periodic signal depicted in \figref{fig:periodic} has $\xi$=0.861 and a period $P$=21.96 hours, and \figref{fig:non-periodic} depicts a non-periodic signal, with $\xi$=0.259 and a period $P$=208.15 hours. Note the spread of frequencies in the DFT, which is
a result of the square signals, and the clear multiplications in the auto-correlation of the periodic signal as opposed to the sporadic signal.

\vspace{-4mm}

\begin{figure}[tbh]
\centering
    \subfloat[Average period per AS]{
    \label{fig:period-per-as}
	\epsfig{file=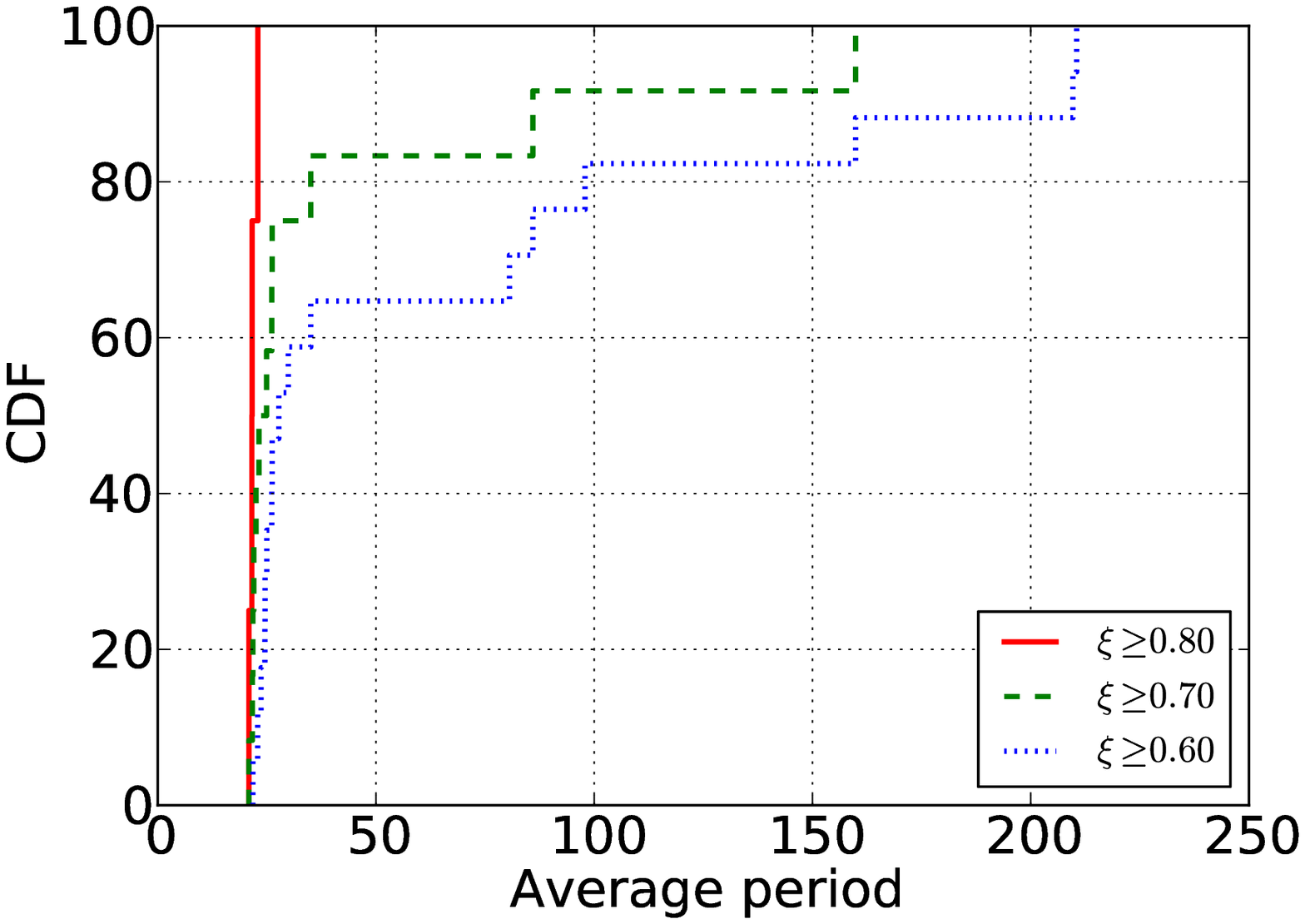,width=0.50\textwidth}
    }
    \hspace{-8mm}
    \subfloat[Period per AS $\left(\xi \geq 0.5\right)$]{
    \label{fig:bars-period-per-as}
	\epsfig{file=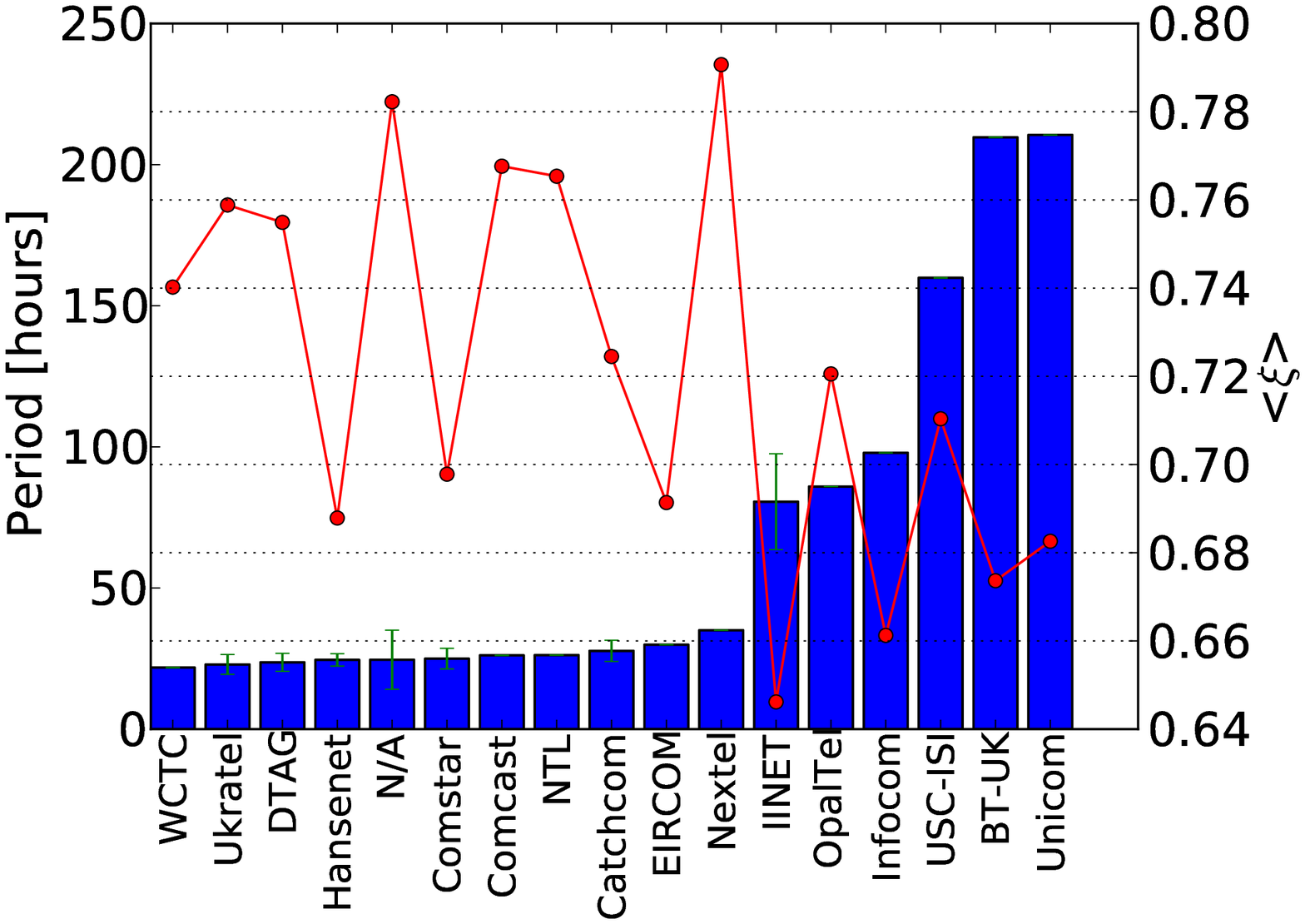,width=0.50\textwidth}
    }
    \caption{CDFs of the average period per AS, and AS details using $\xi geq 0.5$}
    \label{fig:results-period}
\end{figure}



\vspace{-2mm}

\ignore{
\figref{fig:xi-cdf} depicts the distribution of $\xi$, and shows
that almost 80\% of the windows have $\xi>0.5$, which represents a high level of confidence. However,
to be even more conservative, we selected a higher threshold of $\xi>0.6$ and dropped more than 30\% of the online windows. 
}

\figref{fig:results-period} depicts the results of applying the methods on the dataset, while considering 
only online windows that contain
more than 3 intervals, and 
aggregating to the AS-level. \figref{fig:period-per-as} shows the cumulative distribution of the weighted average period per AS for increasing threshold values for $\xi$. The average period per AS is
taken over all agents contained in that AS, weighted by their average $\xi$, which is taken over all online windows per agent.

The figure shows that for all three thresholds, over 60\%
of the ASes have a period that ranges from 20
to 30 hours. As the threshold increases, the average periods become shorter. Surprisingly, the median period is 24 hours, which is similar to the median value
seen in \figref{fig:max-interval}. However, this analysis takes into account the noise, and provides confidence, hence
is significantly more reliable.

\figref{fig:bars-period-per-as} provides a breakdown of the ASes for $\xi \geq 0.6$, showing the average period per AS on the left y-axis
and the average $\xi$ on the right y-axis. Error bars depict the standard deviation from the average period, which is, as can be seen
in the figure, quite small. The figure strengthens the observation that on average, long periods exhibit smaller confidence than
short periods. Furthermore, it shows that a rather small number of ASes actually apply periodic policies, and most of them
are small regional ISPs or ASes owned by small companies. Using a threshold of 0.5, 560 agents in 17 ASes were identified as periodic, while
for a threshold of 0.8, 185 agents remain in only 4 ASes.

\section{Conclusion}
\label{sec:conclusion}
\vspace{-2mm}
This paper presents a measurement study of the dynamics of IP allocation policies in the Internet, which is important
for black-lists accuracy and peer-to-peer applications. Using a two-months
study,
we show that over 60\% of the analyzed hosts have a single IP. Using signal processing methods for overcoming 
inherent noise in the measurements, we found that some of the remaining hosts exhibit periodical patterns of
IP alternation, with a median period of 24 hours.

The findings in this paper present a serious issue with the way blacklists are maintained, as a large portion of the blacklisted IP addresses
might not be valid after as little as 24 hours. Therefore, any service that relies on a prolonged association of routable IP address
and a host must refresh these bindings quite often, or otherwise this matching is most probably wrong.
\\
\\
\ignore{\noindent \textbf{Acknowledgment.} This work was partially funded by the Israeli Science Foundation, grant 1685/07.}

\vspace{-9mm}
%
%
\bibliographystyle{splncs}
\bibliography{visitlog_arxiv}
\end{document}